\documentclass[prd,aps,preprint]{revtex4}

\usepackage{amssymb}
\usepackage{epsfig}
\usepackage{subfigure}
\usepackage{color}
\usepackage{array}
\usepackage{url}
\input epsf.tex
\def\DESepsf(#1 width #2){\epsfxsize=#2 \epsfbox{#1}}

\begin{document}



\title{NEUTRINO MASS AND GRAND UNIFICATION OF FLAVOR}

\author{R. N. MOHAPATRA$^*$ }

\address{Maryland Center for Fundamental Physics, University of Maryland,\\
Maryland, MD-20742, USA\\
$^*$E-mail: rmohapat@umd.edu}

\begin{abstract}
The problem of understanding quark mass and mixing hierarchies has
been an outstanding problem of particle physics for a long time.
The discovery of neutrino masses in the past decade, exhibiting
mixing and mass patterns so very different from the quark sector
has added an extra dimension to this puzzle. This is specially
difficult to understand within the framework of conventional grand
unified theories which are supposed to unify the quarks and
leptons at short distance scales. In the paper, I discuss a recent
proposal by Dutta, Mimura and this author that appears to provide
a promising way to resolve this puzzle. After stating the ansatz,
we show how it can be realized within a SO(10) grand unification
framework. Just as Gell-Mann's suggestion of SU(3) symmetry as a
way to understand the hadronic flavor puzzle of the sixties led to
the foundation of modern particle physics, one could hope that a
satisfactory resolution of the current quark-lepton flavor problem
would provide fundamental insight into the nature of physics
beyond the standard model.
\end{abstract}



\maketitle

\vskip1.0in

\newpage

\baselineskip 18pt

\section{Introduction}
The quark masses as well as their mixings exhibit a hierarchical
pattern i.e. for masses $m_{u,d} \ll m_{c,s} \ll m_{t,b}$; for
mixing angles $V_{ub} \ll V_{cb} \ll V_{cd} \ll V_{ud,cs,tb}$.
This is known as the flavor puzzle for quarks. Unravelling this
puzzle has long been recognized as a challenge for physics beyond
the standard model \cite{weinberg}.In the 1960's, a puzzle of
similar nature was the focus of attention of many when particle
physicists tried to understand why there were different baryons
and mesons with masses close to each other. This puzzle, the
flavor puzzle of the sixties was solved in a seminal paper by
Prof. Gell-Mann- the famous ``Eight-fold Way'' paper, that
practically the first read for every graduate student aspiring to
be a particle theorist in the sixties. In this paper, he proposed
that there is an underlying symmetry for hadrons interactions, the
SU(3) symmetry which is responsible for the closeness of observed
mass and quantum number patterns. This led to the so called
Gell-Mann-Okubo mass formula, which was extremely successful in
understanding the baryon and meson spectra and it culminated with
the discovery of the $\Omega^-$ meson by N. Samios et al. This
proposal of Gell-Mann unleased an idea that had a profound impact
on particle physics: it led to the concepts of quarks as the
constituents of hadrons which forms the foundation of modern
particle physics. It subsequently led to the birth of Quantum
Chromodynamics as the theory of strong forces etc.

In the modern day particle physics of quarks and leptons, the hope
has always been that solving the flavor puzzle may have similar
ground breaking implication for physics beyond the standard model.

The puzzle of quark flavors was already mysterious but it got more
so after the discovery of neutrino masses and mixings in 1998 and
subsequent years. Unlike the quark mixings, lepton mixings do not
exhibit a hierarchical pattern (i.e. neutrino mixings between
generations denoted by $\theta_{ij}$ are given by
$\theta^l_{23}\sim 45^{\rm o}$ and $\theta^l_{12}\simeq 35^{\rm
o}$ as against $\theta^q_{23}\sim 2.5^{\rm o}$ and
$\theta_{12}^q\sim 13^{\rm o}$) and the neutrino masses also do
not exhibit as strong a hierarchy as quarks or charged leptons
i.e. for a normal hierarchy for neutrinos, $m_2/m_3 \simeq V_{us}
\ll m_\mu/m_\tau$. In fact the lepton mixing matrix (the PMNS
matrix) appears to very closely resemble the following pattern:
called the tri-bi-maximal mixing matrix\cite{tbm}:
\begin{equation}
U_{\rm TB} = \left( \begin{array}{ccc}
                 \sqrt{\frac23} & \sqrt{\frac13} & 0 \\
                 -\sqrt{\frac16}  & \sqrt{\frac13}& -\sqrt{\frac12} \\
                 -\sqrt{\frac16} & \sqrt{\frac13} & \sqrt{\frac12}
               \end{array}
        \right).
\end{equation}
On top of this, neutrinos being electrically neutral particles
could be their own anti-particles, the so-called Majorana
fermions. In fact, the most popular way to understand the small
neutrino masses seems to be the seesaw mechanism\cite{seesaw},
which predicts that the neutrinos are their own anti-particles. It
could be that the different mixing pattern for leptons is related
to this feature. Our proposal below has this as one of its
ingredients.

The problem of quark-lepton masses and mixings becomes specially
puzzling in grand unified theories where the quarks and leptons
unify at a very high scale. One would naively expect that when
quarks and leptons unify, their masses and mixings would exhibit a
similar pattern. In fact the seesaw mechanism also suggests that
the scale of neutrino mass is about $10^{14}$ GeV which is very
close to the conventional grand unification scale $\sim 10^{16}$
GeV or so. This then raises the fundamental question of how we
understand the diverse pattern of quarks and leptons in a seesaw
motivated grand unified theory. Cracking this code may provide a
hint of some really new exciting underlying physics.

In this note, I describe a recent ansatz proposed by Dutta and
Mimura and this author which promises a new way to have a unified
understanding of quark lepton flavor\cite{dmm} and its realization
in SO(10) grand unified theories .

 As a prelude to this discussion, let us realize that one way to
 understand the quark mass hierarchy is to start with a rank one
 mass matrix for up, down quarks and charged leptons in the leading order i.e.
 \begin {eqnarray}
 M_{u,d,l}~=~m_{t,b,\tau}\left(\begin{array}{ccc} 0 & 0 & 0 \\0 &
 0 & 0\\0 & 0 & 1\end{array}\right)
 \end{eqnarray}
 and consider corrections coming from non-leading operators.
 Second clue for our ansatz is the observation that small quark
 mixings are an indication that one could write the up and down quark mass
 matrices as a sum a ``big" matrix plus a small matrix with the ``big'' matrix part for
 both sectors being
 proportional to each other so that in the leading order the CKM angles vanish e.g.
 \begin{eqnarray}
 M_{u,d}~=~M^0_{u,d}+\delta_{u,d}
 \end{eqnarray}
 with $M^0_d = r M^0_u$ being the ``big" matrix and $\delta_{u,d}$
 being the smaller part. As just mentioned, the proportionality of the large parts of
 the mass matrices guarantees that the mixing angles will
 necessarily be small since the diagonalizing matrix for the large
 parts are ``parallel" or ``aligned" and the nontrivial CKM
 matrix represents the ``small" misalignment between the two
 matrices determined by the ``smaller" parts of the mass matrix.

 We can therefore now state our ansatz\cite{dmm} which consists of
 two parts:

 \begin{enumerate}

 \item The quark and lepton mass matrices have the following
 general feature:
 \begin{eqnarray}
 M_u~=~M_0~+~\delta_u;\\ \nonumber
M_d~=~rM_0~+~\delta_d;\\ \nonumber M_l~=~rM_0~+~\delta_l;\\
\nonumber M_\nu~=~fv_L
\end{eqnarray}

\item   $M_0$ has rank one.

\end{enumerate}
Note that by a choice of the lepton basis, we can make $f$
diagonal without loss of generality. It is then clear from the
first part of the ansatz (item one) that for an ``anarchic" form
of the matrices $M_0$ and $\delta$ as long as $\delta_{ij} \ll
M_0$, the lepton mixing angles are large whereas the quark mixing
angles are small. The second rank one property than guarantees
that quark and charged lepton masses are hierarchical whereas
since $f$ matrix is arbitrary, any hierarchy in the neutrino
sector is likely to be milder. Incidentally, rank one property to
understand mass hierarchy has been used in the past; see for
instance\cite{rank}.

\section{Gauge group required to implement the ansatz}
The question that arises next is how to implement our ansatz with
a gauge model framework. To implement the first part, it is
important to notice that the up and down quark mass matrices must
be proportional to each other. Such relations do not emerge from
the standard model since the $u_R$ and $d_R$ fields are separate
fields and their Yukawa couplings responsible for the up and down
quark mass matrices are therefore independent of each other. The
situation however changes once we expand the gauge group to the
left-right symmetric group $SU(2)_L\times SU(2)_R\times
U(1)_{B-L}$ group since the $(u_R, d_R)$ form a doublet of the
$SU(2)_R$ group and thus relate the up and down Yukawa
matrices\cite{bdm}. Since quark-lepton unifications arises
naturally within grand unification theories, the obvious group to
consider is the SO(10) group as we do in the next section,
although the basic conditions of the ansatz could also be realized
with less predictive power within the $SU(2)_L\times SU(2)_R\times
SU(4)_c$ partial unification groups.

\section{SO(10) realization of flavor unification ansatz}
As is well known, in SO(10) models, the matter fermions belong to
{\bf 16}-dim. spinor representations. To get fermion masses, we
will consider SO(10) models with {\bf 10}, {\bf 126} plus possibly
another {\bf 10} or {\bf 120} Higgs fields where fermion masses
are generated by renormalizable Yukawa couplings \cite{babu} only
and where type II seesaw\cite{type2} is responsible for neutrino
masses \cite{goran}. To implement our idea, we require that one of
the {\bf 10} Yukawa couplings is the dominant one contributing to
up, down and charged lepton masses and has rank one with other
smaller couplings providing neutrino masses as well as most of the
quark lepton flavor hierarchy. We postpone the discussion of how
to get rank one till later. Let us see if this model does indeed
give us our ansatz. It is well known that in these
models\cite{babu}, we have the following form for all fermion
masses:
\begin{eqnarray}
Y_u &=& h + r_2 f +r_3 h^\prime, \label{eq2} \\\nonumber Y_d &=&
r_1 (h+ f + h^\prime)\,, \\\nonumber Y_e &=& r_1 (h-3f + c_e
h^\prime)\,, \\\nonumber Y_{\nu^D} &=& h-3 r_2 f + c_\nu h^\prime,
\end{eqnarray}
where $Y_a$ are mass matrices divided by the electro-weak vev
$v_{wk}$ and $r_i$ and $c_{e,\nu}$ are the mixing parameters which
relate the $H_{u,d}$ to the doublets in the various GUT
multiplets.
More precisely, the matrices $h$, $f$ and $h^\prime$ in $Y_a$ are
multiplied by the Higgs mixing parameters when they appear in the
fermion mass matrices.

Furthermore, we use the type II seesaw formula\cite{type2} for
getting neutrino masses gives \cite{goran}.
\begin{eqnarray}
{\cal M}_\nu~=~fv_L. \label{eq4}
\end{eqnarray}
Note that $f$ is the same coupling matrix that appears in the
charged fermion masses in Eq. (\ref{eq2}), up to factors from the
Higgs mixings and the Clebsch-Gordan coefficients. This helps us
to
 connect the neutrino parameters to the quark-sector parameters. The equations
(\ref{eq2}) and (\ref{eq4}) are the key equations in our unified
approach to addressing the flavor problem and obviously satisfy
our flavor unification ansatz.

\section{Implementing rank one strategy}

The rank one Yukawa coupling with {\bf 10} Higgs field generates
the features of flavor hierarchy, and rank 1 matrices can often
appear in various ways (flavor symmetry, discrete symmetry, and
string models). In this section, we give an SO(10) model, where
the rank one ansatz used in our discussion of flavor emerges from
extra vector like spinors above the GUT scale as well as a
discrete symmetry.

When the direct couplings of chiral fermions with a Higgs field
are forbidden by the chosen discrete symmetry, and the effective
Yukawa couplings are generated by propagating vector-like matter
fields, the rank of the effective Yukawa matrix depends on the
number of the vector-like fields. Actually, when there are only
one pair of vector-like matter fields as a flavor singlet, the
effective Yukawa matrix is rank 1.

To illustrate this in a warm-up example, we consider a model which
has one extra vector-like pair of matter fields to start with with
mass slightly above the GUT scale contributing to the {\bf 10}
coupling (denoted by
$\psi_V\equiv {\bf 16}_V\oplus \bar\psi_V\equiv \overline{\bf
16}_V$)
and three gauge singlet fields $Y_a$. We add a $Z_4$ discrete
symmetry to the model under which the fields $\psi_a\rightarrow i
\psi_a$, and $Y_a\rightarrow -iY_a$. The {\bf 10}-Higgs field $H$
%
is invariant under this symmetry. The gauge invariant Yukawa
superpotential under this assumption is given by
\begin{eqnarray}
W~=~\psi_V H \lambda\psi_V~+
M_V\psi_V\bar{\psi}_V~+~\bar{\psi}_V\sum_a Y_a\psi_a.
\end{eqnarray}
When we give vevs $\langle Y_a\rangle\neq 0$, $\psi_V$ and
$\psi_a$ are mixed. The heavy vector-like fields, $\bar\psi_V$ and
a linear combination of $\psi_V$ and $\psi_a$ (i.e. $M_V \psi_V+
\sum_a Y_a \psi_a$), and the effective operator below its scale
and at the GUT scale is given by:
\begin{eqnarray}
{\cal L}_{eff}~=~\frac{\lambda}{M_V^2+ \sum_a
Y_a^2}\left[\sum_aY_a\psi_a\right] H \left[\sum_b
Y_b\psi_b\right].
\end{eqnarray}
This gives rise to a rank one $h$ coupling. We note that it does
not contradict the $O(1)$ top Yukawa coupling, when $M_V^2 \sim
\sum_a Y_a^2$ (or $M_V^2 < \sum_a Y_a^2$).

If we let the $\overline{\bf 126}$ Higgs field transform like $-1$
under $Z_4$, it can induce the $f$ coupling with rank three. Our
final model given below builds on this but differs in details e.g.
it has got two vectorlike spinor multiplets instead of one etc.

\section{Making the model predictive}
Simply using the above ansatz in the context of an SO(10) model
with mass relations in Eq. (5), turns out to reproduce the
qualitative features of the quark and lepton spectra quite well.
For example in the context of a two generation model (involving
the second and third generation), this simple ansatz predicts
$V_{cb}\simeq (m_s/m_b + e^{i\sigma} m_c/m_t) \cot\theta$, where
$\theta$ is the atmospheric mixing angle. This relation is in
rough agreement with observations to leading order. In addition,
we have at GUT scale $m_b\sim m_\tau$ as well as $m-\mu\sim -3m_s$
also in rough agreement with observations.

Encouraged by these results, we can be more ambitious and start
using this ansatz in combination of other ideas to make as many
predictions as possible. To this end, we note that in the limit
vanishing Yukawa couplings, the standard model has $[U(3)]^5$
global symmetry. It is therefore quite possible that in the final
understanding of flavor, a subgroup of this large symmetry does
play an important role\cite{discrete}, specially subgroups which
have three dimensional representation to fit three generations. In
order to exploit this observation, one may replace all Yukawa
couplings by flavon fields which transform as three dimensional
representations of a subgroup of $[SU(3)]^5$ and consider the
minima of the flavon theory in flavon space as determining the
values of the Yukawa couplings. It turns out that there are
nontrivial examples where this program is realized. In the second
paper of \cite{dmm}, we presented an $S_4$ subgroup example. Below
I briefly recapitulate this example.

\section{The $SO(10)\times S_4\times Z_n$ model of flavor}
Recall that the $S_4$ group is a 24 element group describing
permutations of four distinct objects and has five irreducible
representations with dimensions ${\bf 3_1\oplus 3_2 \oplus 2
\oplus 1_2 \oplus 1_1}$. The distinction between the
representations with subscripts $1$ and $2$ is that the later
change sign under the transformation of group elements involving
the odd number of permutations of $S_4$.

We assign the three families of {\bf 16}-dim. matter fermions
$\psi$ to ${\bf 3_2}$-dim. representation of $S_4$ and the Higgs
field $H$,
$\bar{\Delta}$ and $H'$ to 
${\bf 1_1, 1_2}$, and ${\bf1_1}$ reps, respectively.
We then choose three 
SO(10) singlet flavons $\phi_i$ transforming as ${\bf 3_2, 3_1,
3_2}$ reps of $S_4$ and one gauge and $S_4$ singlet fields $s_1,
s_2$ transforming as $\bf 1_2$ and $\bf 1_1$
 respectively.
%
%
We further assume that at a scale slightly above the GUT scale,
there are two $S_4$ singlet vectorlike pairs of ${\bf 16\oplus
\overline{16}}$ fields denoted by $\psi_{V}$ and $\bar{\psi}_V$.
In order to get the desired Yukawa couplings naturally from this
high scale theory, we supplement the
$S_4$ group by an $Z_n$ 
group with all the above fields belonging to representations given
in the Table I. The fields and representations to generate the
desired Yukawa couplings. $\omega = e^{i \frac{2\pi}{n}}$.
$\alpha=2+a-b$. In addition to the fields below, there are two
gauge and $S_4$ singlet $1_2,1_1$ fields with $Z_n$ quantum
numbers $\omega^a$ and $\omega^b$ respectively.
\begin{table}
\begin{tabular}{|c|c|c|c|c|c|c|c|c|c|c|c|c|c|c|
} \hline & $\psi$ & $H$ & $\bar\Delta$ & $H^\prime$ & $\phi_1$ &
$\bar\phi_1$ & $\phi_2$ & $\bar\phi_2$ & $\phi_3$ & $\bar\phi_3$ &
$\psi_{V1}$ & $\bar\psi_{V1}$ & $\psi_{V2}$ & $\bar\psi_{V2}$
\\ \hline
SO(10) & $\bf 16$ & $\bf 10$ & $\overline{\bf 126}$ & $\bf 10$ &
$\bf 1$ & $\bf 1$ & $\bf 1$ & $\bf 1$ & $\bf 1$ & $\bf 1$ & $\bf
16$ & $\overline{\bf 16}$ & $\bf 16$ & $\overline{\bf 16}$
\\ \hline
$S_4$ & $\bf 3_2$ & $\bf 1_1$ & $\bf 1_2$ & $\bf 1_1$ & $\bf 3_2$
& $\bf 3_2$ & $\bf 3_1$ & $\bf 3_1$ & $\bf 3_2$ & $\bf 3_2$ & $\bf
1_1$ & $\bf 1_1$ & $\bf 1_2$ & $\bf 1_2$
\\ \hline
%
%
$Z_n$ & $1$ & $\omega^{-4}$ & $\omega^{-2-a}$ & $\omega^{-1}$ &
$\omega^{2}$ & $\omega^{-2}$ & $\omega$ & $\omega^{-1}$ &
$\omega^\alpha$ & $\omega^{-\alpha}$ & $\omega^2$ & $\omega^{-2}$
& $\omega$ & $\omega^{-1}$
\\ \hline
\end{tabular}
\end{table}

The most general high scale  Yukawa superpotential involving
matter fields invariant under this symmetry is given by:
\begin{eqnarray}\label{generalYukawa}
W = (\phi_1 \psi)\bar{\psi}_{V1}~+~ \psi_{V1} \psi_{V1} H ~+~M_1
\bar{\psi}_{V1} \psi_{V1}~\\ \nonumber +(\phi_2
\psi)\bar{\psi}_{V2}~+~ \frac{1}{M_P}s_1 \psi_{V2} \psi_{V2}
\bar\Delta +~M_2\bar{\psi}_{V2} \psi_{V2}\\ \nonumber ~+
\frac{1}{M^2_P}s_2 (\phi_3 \psi \psi) \bar \Delta +
\frac{1}{M_P}(\phi_2 \psi \psi) H^\prime,
\end{eqnarray}
where the brackets stand for the $S_4$ singlet contraction of
flavor index. The singlet field $s_i$ can have large vev as
follows: consider its $Z_n$ charge to be such that the only
polynomial term involving the $s_i$ in the superpotential has the
form $s_i^{k_i}/M^{k_i-3}_{P}$ (in order to describe the essential
potential, we ignore a possible $s_1^{\ell_1} s_2^{\ell_2}$ term).
The dominant part of the potential in the presence of SUSY
breaking has the form:
\begin{eqnarray}
V(s_i)~=~-m^2_{s_i}
|s_i|^2+k\frac{s_i^{2k_i-2}}{M^{2k_i-6}_P}+\cdots.
\end{eqnarray}
Minimizing this leads to $\langle s_i
\rangle\sim[m^2_{S_i}M^{2k_i-6}_P]^{\frac{1}{2k_i-4}}$, which is
above GUT scale for larger values of the integer $k_i$ (which in
turn is determined by the $Z_n$ symmetry charge of $s_i$).
One could also have large vevs for $s_1, s_2$ by using anomalous
$U(1)$ charges for them using $D$-terms to break the $U(1)$
symmetry.

The effective theory below the scales $M_{1,2}$ and $\langle s_i
\rangle$ of the vector-like pair masses and the $s_i$-vevs
respectively is given by:
\begin{eqnarray}
W = (\phi_1 \psi) (\phi_1 \psi) H + (\phi_2 \psi) (\phi_2 \psi)
\bar\Delta + (\phi_3 \psi \psi) \bar \Delta + (\phi_2 \psi \psi)
H^\prime, \label{phiyukawa}
\end{eqnarray}
where we have omitted the dimensional coupling constants to make
it simple for the purpose of writing. The discrete symmetries
prevent $\phi^2/M^2$ corrections to these terms. So our
predictions based on this effective superpotential do not receive
large corrections.
We note that the non-renormalizable terms in
Eq.(\ref{generalYukawa}) can also be obtained from renormalizable
couplings if we introduce further $S_4$-triplet vectorlike fields.
Here, however we use only $S_4$-singlet vectorlike fields to get
rank 1 contribution to $h$ and $f$ Yukawa couplings and that is
why we need the non-renormalizable terms to be present in
Eq.(\ref{generalYukawa}.

In order to get  fermion masses, we have to find the alignment
\cite{Ross} of the vevs of the flavon fields $\phi_{1,2,3}$. We
show below that the following choice of vevs are among the minima
of the flavon superpotential provided the couplings of mixed terms
between different $\phi_i$'s are small compared to other
couplings:
%
%
%
\begin{equation}
\phi_1 = \left(
\begin{array}{c}
0 \\ 0 \\ 1
\end{array}
\right), \quad \phi_2 = \left(
\begin{array}{c}
0 \\ -1 \\ 1
\end{array}
\right), \quad \phi_3 = \left(
\begin{array}{c}
1 \\ 1 \\ 1
\end{array}
\right). \label{flavon}
\end{equation}
Clearly, there are other vacua for the flavon model that we do not
choose. What is however nontrivial is that the alignments are
along quantized directions. This is a consequence of supersymmetry
combined with discrete symmetries in the theory. Given these vev,
we find from Eq. (\ref{phiyukawa}) that the Yukawa coupling
matrices $h,f,h'$ have the form:
\begin{eqnarray}
h &\propto& \left(
\begin{array}{ccc}
0 & 0 & 0 \\
0 & 0 & 0 \\
0 & 0 & 1
\end{array}
\right),
\\
f &\propto& \left(
\begin{array}{ccc}
0 & 0 & 0 \\
0 & 1 & -1 \\
0 & -1 & 1
\end{array}
\right)
 + \lambda
\left(
\begin{array}{ccc}
0 & 1 & 1 \\
1 & 0 & 1 \\
1 & 1 & 0
\end{array}
\right), \\
h^\prime &\propto& \left(
\begin{array}{ccc}
0 & 1 & -1 \\
1 & 0 & 0 \\
-1 & 0 & 0
\end{array}
\right),
\end{eqnarray}
and the charged fermion mass matrices can then be inferred.
 The neutrino mass matrix in this basis has the form:
\begin{equation}
{\cal M}_\nu~=~\left(\begin{array}{ccc} 0 & c & c\\c & a & c-a \\
c & c-a & a\end{array}\right),
\end{equation}
where $c/a = \lambda \ll 1$. It is diagonalized by the
tri-bi-maximal matrix
\begin{equation}
U_{\rm TB} = \left( \begin{array}{ccc}
                 \sqrt{\frac23} & \sqrt{\frac13} & 0 \\
                 -\sqrt{\frac16}  & \sqrt{\frac13}& -\sqrt{\frac12} \\
                 -\sqrt{\frac16} & \sqrt{\frac13} & \sqrt{\frac12}
               \end{array}
        \right).
\end{equation}
 This is however not the full PMNS matrix which will
receive small corrections from diagonalization of the charged
lepton matrix, which not only make small contributions to the
$\theta_{\rm atm}$ and $\theta_{\odot}$ but also generate a small
$\theta_{13}$.

The neutrino masses are given by $m_{\nu3}~=~2a-c\,$;
$m_{\nu2}~=~2c$ and $m_{\nu1}~=~-c$. To fit observations, we
require $\lambda = c/a \simeq \sqrt{\Delta m_{\odot}^2/\Delta
m_{\rm atm}^2} \sim 0.2$, which fixes the neutrino masses
$m_{\nu3} \simeq 0.05$ eV, $m_{\nu2} \simeq 0.01$ eV, and
$m_{\nu1} \simeq 0.005$ eV. We will see below that $\lambda$ is
also the Cabibbo angle substantiating our claim that neutrino mass
ratio and Cabibbo angle are related.


 For the charged lepton, up and down quark mass matrices, we have:
\begin{eqnarray}
M_\ell ~=~\frac{r_1}{\tan\beta}\left(\begin{array}{ccc} 0 & -3m_1
+ \delta &
-3m_1-\delta \\ -3m_1 + \delta & -3m_0 & 3m_0-3m_1 \\
-3m_1-\delta & 3m_0-3m_1 & -3m_0+M\end{array}\right),\\ \nonumber
M_d ~=~\frac{r_1}{\tan\beta}\left(\begin{array}{ccc} 0 & m_1 +
\delta &
m_1-\delta \\ m_1 + \delta & m_0 & -m_0+m_1 \\
m_1-\delta & -m_0+m_1 & m_0+M\end{array}\right),\\ \nonumber M_u
~=~\left(\begin{array}{ccc} 0 & r_2m_1 + r_3\delta &
r_2m_1-r_3\delta \\ r_2m_1 + r_3 \delta & r_2m_0 & -r_2m_0+r_3m_1 \\
r_2m_1+r_3\delta & -r_2m_0+r_3m_1 & r_2m_0+M\end{array}\right),
\end{eqnarray}
where $\tan\beta$ is a ratio of $H_{u,d}$ vevs. Note that $m_1/m_0
= \lambda \sim 0.2$ and of course $m_0 \ll M$. A quick examination
of these mass matrices leads to several immediate conclusions:

\begin{enumerate}

\item The model predicts that at GUT scale $m_b\simeq m_\tau$.

\item Since $(M_d)_{11} \to 0$, we get $V_{us} \simeq
\sqrt{m_d/m_s}$.

\item The empirically satisfied relation $m_\mu m_e\simeq m_s m_d$
can be obtained by the choice of parameters $ -3m_1+\delta  =
(m_1+\delta)e^{i\sigma}$, where $\sigma$ is a phase. Solving this
equation, we find that $\delta = m_1 (1+i \cot \sigma/2)$. We
obtain $V_{us} \simeq (1-r_3/r_2) \delta/m_0$, thereby relating
Cabibbo angle to the neutrino mass ratio $m_{\odot}/m_{\rm
atm}\simeq \lambda$.

\item $m_\mu\sim -3 m_s$.

\item The leptonic mixing angle to diagonalize $M_\ell$ is related
to quark mixing $\theta^l_{12}\sim \frac13
 V_{us}$, which leads to a prediction for $\sin\theta_{13}\equiv
 U_{e3}\sim \frac{V_{us}}{3\sqrt{2}}\simeq 0.05$ \cite{king}.

 \item  $\displaystyle V_{cb}\sim \frac{m_s}{m_b} \cot \theta_{\rm atm}$.

\item The masses of up and charm quarks are given by the
parameters $r_{2,3}$ and are therefore not predictions of the
model.

\item CP violation in quark sector can put in by making the
parameters $h'$ complex.

\item The model predicts a small amplitude for neutrino-less
double beta decay from light neutrino mass: $m_{\nu_{ee}} \sim c
\sin\theta_{12}^l \simeq 0.3$ meV.

 \end{enumerate}

 The first four relations are fairly well satisfied by
 observations; the fifth  prediction (i.e. that for $U_{e3}$)
  can be tested in upcoming reactor and
  long baseline experiments. Note that the deviation from
  tri-bi-maximal mixing pattern coming from the charged lepton
  mass diagonalization could be thought of as a small perturbation of
  the neutrino mass matrix except that we predict
  the form of the perturbation from symmetry considerations.
The sixth prediction gives a smaller value for $V_{cb}$ ($0.02$ as
against observed GUT scale value of $0.03$) if one uses GUT scale
extrapolated value of the known $b$ mass. However, in the MSSM
there are threshold corrections to the $b-s$ quark mass mixing
from gluino and wino exchange one-loop diagrams; by choosing this
contribution, one could obtain the desired $V_{cb}$.

Note that in this model, the top quark Yukawa coupling at GUT
scale
 arises from an effective higher dimensional operator.
We have showed the effective operator in Eq.(\ref{phiyukawa}) by
expanding $\phi/M$. The more precise form for the top Yukawa
coupling is $\phi^2/(M_1^2+\phi^2) h_{\psi_V \psi_V H}$, where
$h_{\psi_V \psi_V H}$ is a coupling of $\psi_V \psi_V H$ term, and
$\phi$ is the vev of $\phi_1$ multiplied by $\phi_1 \psi
\bar\psi_V$ coupling. This is simply because the low energy third
generation field is a linear combination of the form $\cos
\alpha\, \psi_3- \sin \alpha\, \psi_V$  with the mixing angle
$\sin \alpha \simeq \phi/\sqrt{M_1^2+\phi^2}$.

 Therefore, in general,
there is no gross contradiction to the fact that the top Yukawa
coupling is order 1. However, in our case, if $\phi/M_1$ becomes
close to 1, the atmospheric mixing shifts from the maximal angle.
Given the error in the determination of the atmospheric mixing
angle, this is consistent with data and as this measurement
sharpens, this is going to provide a test of this particular
model.
%
%
 The desired smallness of the effective $f$ and $h^\prime$ couplings
however are more naturally obtained due to the presence of the
Planck mass in the denominator. In order to make the $f$-coupling
dominate over the $h^\prime$, we have to choose a small coupling
for the $H^\prime$ Higgs field in Eq. (4). Similarly the $\lambda$
term in Eq. (9) is assumed to be small compared to the coefficient
of the first matrix.

Thus within these set of assumptions, this model is in good
phenomenological agreement with observations. In a more complete
theory, these assumptions need to be addressed. We however find it
remarkable that despite these shortcomings, the model provides a
very useful unification strategy of the diverse quark-lepton
mixing patterns.

\section{Vev alignment as minima of flavon theories}
In this section, we give examples of how the minima of flavon
theories can determine the Yukawa couplings of the fermions and
lead to predictive flavor models. We discuss the specific case of
the $S_4$ model at hand. This mechanism is of course applicable to
any general group.

 We start our discussion by
giving some simple examples and discussing the flavon alignment as
a prelude to the more realistic example. First thing to note is
that ${\bf 3_1}^3$ is invariant under $S_4$, but ${\bf 3_2}^3$ is
not. Denoting $\phi = (x,y,z)$, we see that in the first case, the
singlet of $\phi^3 = xyz$. The superpotential for a $\bf 3_1$
flavon field $\phi$ can therefore be written as
\begin{equation}
W= \frac12 m \phi^2 - \lambda \phi^3 = \frac12m(x^2+y^2+z^2) -
\lambda xyz.
\end{equation}
The solution of $F$-flat vacua ($\phi\neq 0$) are
\begin{equation}
\phi = \frac{m}{\lambda} \{(1,1,1)\ {\rm or}\ (1,-1,-1)\ {\rm or}\
(-1,1,-1)\ {\rm or}\ (-1,-1,1)\}.
\end{equation}
 These vacua break $S_4$ down to $S_3$ and in the process determine the Yukawa couplings.

On the other hand, when $\bf 3_2$ flavon is used (or the cubic
term is forbidden by a discrete symmetry), quartic term involving
the triplet is crucial for the $F$-flat vacua.
The invariant quartic term $\phi^4$ gives two linear combinations
of the form $x^4+y^4+z^4$ and $x^2 y^2+ y^2 z^2+z^2 x^2$. This is
because they have to be symmetric homogenous terms and invariant
under the Klein's group, which is $\pi$ rotation around the
$x,y,z$ axes.

Thus, the superpotential term for $\bf 3_2$ field $\phi$ is
\begin{eqnarray}
W &=& \frac12 m \phi^2 - \frac{\kappa^{(1)}}{M} (\phi^4)_1 -
\frac{\kappa^{(2)}}{M} (\phi^4)_2 \\
&=& \frac12(x^2+y^2+z^2)- \frac{\kappa^{(1)}}{4M} (x^4+y^4+z^4) -
\frac{\kappa^{(2)}}{2M}(x^2y^2+y^2z^2+z^2x^2). \nonumber
\end{eqnarray}
The nontrivial $F$-flat vacua ($\phi\neq 0$) are
\begin{equation}
\phi = \sqrt{\frac{mM}{\kappa^{(1)}}}\, \vec{a},\quad
       \sqrt{\frac{mM}{\kappa^{(1)}+2\kappa^{(2)}}}\, \vec{b}, \quad
       \sqrt{\frac{mM}{\kappa^{(1)}+\kappa^{(2)}}}\, \vec{c},
\label{the-vacua}
\end{equation}
where $\vec{a} = (0,0,\pm1)$, $(0,\pm1,0)$, $(\pm1,0,0)$, $\vec{b}
= (\pm1,\pm1,\pm1)$, and $\vec{c} = (0,\pm1,\pm1)$,
$(\pm1,\pm1,0)$, $(\pm1,0,\pm1)$. We note that these vectors
correspond to the axes of the regular hexahedron. The vacua break
$S_4$ down to $Z_4$, $Z_3$, and $Z_2$, respectively. More
importantly, the vacuum states in Eq. (12) used in the analysis of
fermion masses in the previous section are a subset of the above
vacua.
%

Note that if we add a $\phi^4$ term to the superpotential
involving the $\bf 3_1$ flavon field, $\vec{a}$ vacuum is
possible, in addition to the original $\vec{b}$ vacua.
However, $\vec{c}$ vacuum is absent. 

\section{Comments}

A complete understanding of flavor is clearly a very ambitious
task. Our proposal should be considered as a simple beginning
towards a final theory. It should be noted that even though we
have considered on SO(10) group, our general unification ansatz
(without as much predictivity) in $SU(2)_L\times SU(2)_R\times
SU(4)_c$ theories as well and perhaps other groups such as $E_6$.
Similarly, one should explore other flavor models.

A second point of importance is that while we have kept only
leading order terms, one should clearly consider higher order
corrections to our predictions systematically. In the above mode,
we have checked next order corrections and found them to be absent
due to the discrete symmetries.

A final source of corrections could come from anomalies in the
discrete symmetries, although we expect them to be
small\cite{ratz}.

\section{Conclusion}
In summary, I have discussed a recently proposed ansatz that has
the potential to provide a unified description of the diverse
quark and lepton flavor. This could provide the first opening into
a very difficult problem of particle physics- the problem of
flavor. A simple realization of this ansatz is shown to occur
within a grand unified SO(10) model with type II seesaw describing
the neutrino masses. The successes of that model are that it seems
to provide an understanding of several observed quark-lepton mass
relations such as bottom-tau mass unification, strange quark-muon
mass ratio ($1/3$) etc. and predicts a value for $\theta_{13} \sim
0.05$ and atmospheric mixing angle different from the maximal
value. The model like most grand unified theories of neutrinos
predicts a normal hierarchy and observation of inverted hierarchy
will therefore rule out this model (as well as most grand unified
theories). Under certain reasonable approximations, this also
seems to explain why $m_{solar}/m_{atm}\sim \theta_C$. Both the
predictions given above ($\theta_{13}$ and $\theta_{atm}$ ) could
be used to test the model in the upcoming long baseline neutrino
experiments. It also predicts a value of 0.3 meV for the
neutrinoless double beta decay experiments.

\section*{Acknowledgement}
I would like to thank Harald Fritzsch for inviting me to speak at
this stimulating conference honoring Prof. M. Gell-Mann. This work
is supported by the US National Science Foundation under grant No.
PHY-0652363. The author is grateful to Y. Mimura and B. Dutta in
whose collaboration this work was done and to Michael Ratz for
discussions on anomalous discrete symmetries.
%




\begin{thebibliography}{9}
\bibitem{weinberg}
  S.~Weinberg,
  Trans.\ New York Acad.\ Sci.\  {\bf 38}, 185 (1977);
%
  F.~Wilczek and A.~Zee,
  Phys.\ Lett.\  B {\bf 70}, 418 (1977)
  [Erratum-ibid.\  {\bf 72B}, 504 (1978)];
%
  H.~Fritzsch,
  Phys.\ Lett.\  B {\bf 73}, 317 (1978).


\bibitem{tbm}
  P.~F.~Harrison, D.~H.~Perkins and W.~G.~Scott,
  Phys.\ Lett.\  B {\bf 530}, 167 (2002)
  [hep-ph/0202074];
 X.~G.~He and A.~Zee,
  Mod.\ Phys.\ Lett.\  A {\bf 22}, 2107 (2007)
  [arXiv:hep-ph/0702133];
  Z.~z.~Xing,
  Phys.\ Lett.\  B {\bf 533}, 85 (2002)
  [hep-ph/0204049].

\bibitem{seesaw} P.~Minkowski,
Phys.\ Lett.\ B {\bf 67} (1977) 421;
%
T.~Yanagida in {\em Workshop on Unified Theories, KEK Report
79-18}, p.~95, 1979;
%
M.~Gell-Mann, P.~Ramond and R.~Slansky, {\em Supergravity},
p.~315.
\newblock Amsterdam: North Holland, 1979;
%
S.~L. Glashow, {\em 1979 Cargese Summer Institute on Quarks and
Leptons}, p.~687.
\newblock New York: Plenum, 1980;
%
R.~N. Mohapatra and G.~Senjanovic,
 Phys.\ Rev.\ Lett,\ {\bf 44} (1980) 912.



\bibitem{dmm} B.~Dutta, Y.~Mimura and R.~N.~Mohapatra,
  Phys.\ Rev.\  D {\bf 80}, 095021 (2009); JHEP {\bf 1005}, 034
  (2010).

  \bibitem{rank} B.~S.~Balakrishna, A.~L.~Kagan and R.~N.~Mohapatra,
  Phys.\ Lett.\  B {\bf 205}, 345 (1988);  K.~S.~Babu and R.~N.~Mohapatra,
  Phys.\ Rev.\ Lett.\  {\bf 64}, 2747 (1990); B.~A.~Dobrescu and P.~J.~Fox,
  JHEP {\bf 0808}, 100 (2008);  L.~Ferretti, S.~F.~King and A.~Romanino,
  JHEP {\bf 0611}, 078 (2006).

  \bibitem{bdm} K.~S.~Babu, B.~Dutta and R.~N.~Mohapatra,
  Phys.\ Rev.\  D {\bf 60}, 095004 (1999).

  \bibitem{babu}  K.~S.~Babu and R.~N.~Mohapatra,
  Phys.\ Rev.\ Lett.\  {\bf 70}, 2845 (1993).



\bibitem{type2}
  G.~Lazarides, Q.~Shafi and C.~Wetterich,
  Nucl.\ Phys.\  B {\bf 181}, 287 (1981);
%
  J.~Schechter and J.~W.~F.~Valle,
  Phys.\ Rev.\  D {\bf 22}, 2227 (1980);
%
  R.~N.~Mohapatra and G.~Senjanovic,
  Phys.\ Rev.\  D {\bf 23}, 165 (1981).

  \bibitem{goran}
  B.~Bajc, G.~Senjanovic and F.~Vissani,
  Phys.\ Rev.\ Lett.\  {\bf 90}, 051802 (2003).


\bibitem{discrete}
G.~Altarelli and F.~Feruglio,
  Nucl.\ Phys.\  B {\bf 741}, 215 (2006)
  [hep-ph/0512103];
%
  G.~Altarelli, F.~Feruglio and C.~Hagedorn,
  JHEP {\bf 0803}, 052 (2008)
  [arXiv:0802.0090 [hep-ph]];
%
  F.~Feruglio, C.~Hagedorn and L.~Merlo,
  JHEP {\bf 1003}, 084 (2010)
  [arXiv:0910.4058 [hep-ph]];
%
  S.~Morisi and E.~Peinado,
  Phys.\ Rev.\  D {\bf 80}, 113011 (2009)
  [arXiv:0910.4389 [hep-ph]];
%
  M.~C.~Chen and S.~F.~King,
  JHEP {\bf 0906}, 072 (2009)
  [arXiv:0903.0125 [hep-ph]];
%
%
%
  Y.~Cai and H.~B.~Yu,
  Phys.\ Rev.\  D {\bf 74}, 115005 (2006)
  [hep-ph/0608022];
%
  F.~Bazzocchi and S.~Morisi,
  Phys.\ Rev.\  D {\bf 80}, 096005 (2009)
  [arXiv:0811.0345 [hep-ph]];
%
  H.~Ishimori, Y.~Shimizu and M.~Tanimoto,
  Prog.\ Theor.\ Phys.\  {\bf 121}, 769 (2009)
  [arXiv:0812.5031 [hep-ph]];
%
  F.~Bazzocchi, L.~Merlo and S.~Morisi,
  Phys.\ Rev.\  D {\bf 80}, 053003 (2009)
  [arXiv:0902.2849 [hep-ph]];
%
  G.~Altarelli, F.~Feruglio and L.~Merlo,
  JHEP {\bf 0905}, 020 (2009)
  [arXiv:0903.1940 [hep-ph]];
%
  W.~Grimus, L.~Lavoura and P.~O.~Ludl,
  J.\ Phys.\ G {\bf 36}, 115007 (2009)
  [arXiv:0906.2689 [hep-ph]];
%
  G.~J.~Ding,
  Nucl.\ Phys.\  B {\bf 827}, 82 (2010)
  [arXiv:0909.2210 [hep-ph]];
%
  Y.~Daikoku and H.~Okada,
  arXiv:0910.3370 [hep-ph];
%
  M.~K.~Parida,
  Phys.\ Rev.\  D {\bf 78}, 053004 (2008)
  [arXiv:0804.4571 [hep-ph]];
%
%
  C.~Luhn, S.~Nasri and P.~Ramond,
  Phys.\ Lett.\  B {\bf 652}, 27 (2007)
  [arXiv:0706.2341 [hep-ph]];
%
  C.~Hagedorn, M.~A.~Schmidt and A.~Y.~Smirnov,
  Phys.\ Rev.\  D {\bf 79}, 036002 (2009)
  [arXiv:0811.2955 [hep-ph]];
%
 F.~Bazzocchi and I.~de Medeiros Varzielas,
  Phys.\ Rev.\  D {\bf 79}, 093001 (2009)
  [arXiv:0902.3250 [hep-ph]];
  %
  A.~Adulpravitchai, A.~Blum and C.~Hagedorn,
  JHEP {\bf 0903}, 046 (2009)
  [arXiv:0812.3799 [hep-ph]];
%
  S.~F.~King and C.~Luhn,
  Nucl.\ Phys.\  B {\bf 820}, 269 (2009)
  [arXiv:0905.1686 [hep-ph]];
%
%
  M.~C.~Chen and K.~T.~Mahanthappa,
  Phys.\ Lett.\  B {\bf 681}, 444 (2009)
  [arXiv:0904.1721 [hep-ph]];
%
  arXiv:0910.5467 [hep-ph];
%
  For some earlier models, see
  K.~S.~Babu, E.~Ma and J.~W.~F.~Valle,
  Phys.\ Lett.\  B {\bf 552}, 207 (2003)
  [hep-ph/0206292];
%
  K.~S.~Babu and X.~G.~He,
  hep-ph/0507217;
%
  R.~N.~Mohapatra, S.~Nasri and H.~B.~Yu,
  Phys.\ Lett.\  B {\bf 639}, 318 (2006)
  [hep-ph/0605020].

For recent reviews, see G.~Altarelli and F.~Feruglio,
  arXiv:1002.0211 [hep-ph]; H.~Ishimori, T.~Kobayashi, H.~Ohki, H.~Okada, Y.~Shimizu and M.~Tanimoto,
  arXiv:1003.3552 [hep-th].






\bibitem{king}
J.~Ferrandis and S.~Pakvasa, Phys.\ Lett.\  B {\bf 603}, 184
(2004)
   [hep-ph/0409204];  S.~F.~King, JHEP {\bf 0508}, 105 (2005)
   [hep-ph/0506297];
  B.~Dutta and Y.~Mimura,
  Phys.\ Lett.\  B {\bf 633}, 761 (2006)
  [hep-ph/0512171];
%
M.~C.~Chen and K.~T.~Mahanthappa, Phys.\ Lett.\  B {\bf 652}, 34
(2007).


\bibitem{Ross}
  S.~F.~King and G.~G.~Ross,
  Phys.\ Lett.\  B {\bf 520}, 243 (2001)
  [hep-ph/0108112];
%
  S.~F.~King,
  JHEP {\bf 0508}, 105 (2005)
  [hep-ph/0506297];
%
  I.~de Medeiros Varzielas and G.~G.~Ross,
  Nucl.\ Phys.\  B {\bf 733}, 31 (2006)
  [hep-ph/0507176];
%
  I.~de Medeiros Varzielas, S.~F.~King and G.~G.~Ross,
  Phys.\ Lett.\  B {\bf 644}, 153 (2007)
  [hep-ph/0512313];
%
  Phys.\ Lett.\  B {\bf 648}, 201 (2007)
  [hep-ph/0607045].
   S.~F.~King and C.~Luhn,
  Nucl.\ Phys.\  B {\bf 832}, 414 (2010);
  C.~Hagedorn, S.~F.~King and C.~Luhn,
  JHEP {\bf 1006}, 048 (2010).

\bibitem{ratz} M. Ratz, private communications. R.~Kappl, H.~P.~Nilles, S.~Ramos-Sanchez,
M.~Ratz, K.~Schmidt-Hoberg and P.~K.~S.~Vaudrevange,
  Phys.\ Rev.\ Lett.\  {\bf 102}, 121602 (2009)






\end{thebibliography}
\end{document}